\documentclass[twocolumn,prb,aps,floatfix]{revtex4}

\usepackage{graphicx}
\usepackage{dcolumn}
\usepackage{bm}

\begin{document}

\preprint{}

\title{Nominally forbidden transitions in the interband optical spectrum of quantum dots}

\author{Gustavo A. Narvaez}
\affiliation{National Renewable Energy Laboratory, Golden, Colorado 80401}
\author{Alex Zunger}
\affiliation{National Renewable Energy Laboratory, Golden, Colorado 80401}

\date{July 18, 2006}

\begin{abstract}
  We calculate the excitonic optical absorption spectra of (In,Ga)As/GaAs
  self-assembled quantum dots by adopting an atomistic pseudopotential
  approach to the single-particle problem followed by a
  configuration-interaction approach to the many-body problem.  We find three
  types of {\em allowed} transitions that would be naively expected to be
  forbidden.(i) Transitions that are parity forbidden in simple effective mass
  models with infinite confining wells (e.g.  $1S$-$2S$, $1P$-$2P$) but are
  possible by finite band-offsets and orbital-mixing effects; (ii)
  light-hole--to--conduction transitions, enabled by the confinement of
  light-hole states; and (iii) transitions that show and enhanced intensity
  due to electron-hole configuration mixing with allowed transitions.  We
  compare these predictions with results of 8-band ${\bf k}\cdot{\bf p}$
  calculations as well as recent spectroscopic data. Transitions in (i) and
  (ii) explain recently observed satellites of the allowed $P$-$P$
  transitions.
\end{abstract}

\maketitle

\section{Introduction}
%
%
%
%
In quantum dot spectroscopy, rather simple, idealized theoretical approaches
have been applied to discuss which confined interband optical transitions are
formally allowed and which are formally forbidden. But one {\em expects} the
simple rules not to work. Yet, the mechanisms for failure have only been
assessed within extensions of the simple models. To understand these
mechanisms demands a high-level approach that naturally includes the
complexity of the dots. Such approaches to the calculation of the optical
properties are rare, with 8-band ${\bf k}\cdot{\bf p}$ calculations being
among the most sophisticate approaches used so far.
Here, we discuss the nature of confined transitions in lens-shaped
(In,Ga)As/GaAs quantum dots by using an atomistic pseudopotential-based
approach.\cite{zunger_pssb_2001,williamson_PRB_2000,wang_PRB_1999}
Specifically, we study three mechanisms that render nominally forbidden
transitions, in lower approximations, to allowed transitions within more
realistic approximations: (i) ``$2S$-to-$1S$'' and ``$2P$-to-$1P$''
``crossed'' transitions allowed by finite band-offset effects and orbital
mixing; (ii) transitions involving mixed heavy-hole and light-hole states,
enabled by the confinement of light-hole states; and (iii) many-body
configuration-mixing intensity enhancement enabled by electron-hole Coulomb
interaction.
We also compare our results with those of 8-band ${\bf k}\cdot{\bf p}$ calculations. 
Our atomistic pseudopotential theory explains recent spectroscopic data.

\section{Interband optical spectrum of (In,Ga)As/GaAs dots}

\subsection{Method of calculation}
%
%
%
%
In our approach, the atomistic single-particle energies ${\cal E}_i$ and wave
functions $\psi_i$ are solutions to the atomistic Schr\"odinger
equation\cite{zunger_pssb_2001}

\begin{equation}
\label{schrodinger}
\{-\frac{1}{2}\nabla^2+V_{SO}+\sum_{l,\alpha}\,v_{\alpha}({\bf r}-{\bf
  R}_{\,l,\alpha})\}\psi_i={\cal E}_{i}\,\psi_i ,
\end{equation}

\noindent where $v_{\alpha}$ is a pseudopotential for atom of type $\alpha$,
with $l$-th site position ${\bf R}_{\,l,\alpha}$ in either the dot or the GaAs
matrix. These positions are relaxed by minimizing the total elastic energy
consisting of bond-bending plus bond-stretching terms via a valence force
field functional.\cite{williamson_PRB_2000} This results in a realistic strain
profile $\tilde\varepsilon({\bf R})$ in the
nanostructure.\cite{pryor_JAP_1998}
In addition, $v_{\alpha}$ depends explicitly on the isotropic component of the
strain ${\rm Tr}[\tilde\varepsilon({\bf R})]$.\cite{kim_PRB_2002} $V_{SO}$ is
a non-local (pseudo) potential that accounts for spin-orbit
coupling.\cite{williamson_PRB_2000} In the single-particle approximation, the
transition intensity for light polarized along $\hat{\bf e}$ is

\begin{equation}
\label{absorption.SP}
I^{(SP)}(\omega;\hat{\bf e})=\sum_{i,j}\,
|\langle\psi^{(e)}_i|\hat{\bf e}\cdot{\bf p}|\psi^{(h)}_{j}\rangle|^2\big]\delta[\hbar\omega-{\cal E}^{(e)}_{i}+{\cal E}^{(h)}_{j}],
\end{equation}

\noindent where ${\bf p}$ is the electron momentum.\cite{cardona_book}

In addition to the single-particle effects, many-particle effects cause each
of the monoexciton states $\Psi^{(\nu)}(X^0)$ to be a mixture of several
electron-hole pair configurations (Slater determinants) $e_ih_j$. Namely,

\begin{equation}
\label{X0.states}
|\Psi^{(\nu)}(X^0)\rangle=\sum_{i,j}\,C^{\,(\nu)}_{i,j}|e_ih_j\rangle.
\end{equation} 

\noindent The coefficients $C^{\,(\nu)}_{i,j}$ are determined by
the degree of configuration mixing allowed by the electron-hole Coulomb and
exchange interaction.\cite{franceschetti_PRB_1999} This mixing is determined
by the symmetry of the $e$-$h$ orbitals and by their single-particle energy
separation. The many-body optical absorption for (incoherent) unpolarized
light\cite{note_unpolarized} is given by

\begin{equation}
\label{absorption}
I^{(MP)}(\hbar\omega)=\frac{1}{3}\sum_{\nu}\,\sum_{\hat{\bf e}=\hat{\bf
    x},\hat{\bf y},\hat{\bf z}} |M(\hat{\bf e})|^2\,\delta[\hbar\omega-E^{(\nu)}(X^0)],
\end{equation}

\noindent where $M(\hat{\bf e})=\langle\Psi^{(\nu)}(X^0)|\hat{\bf e}\cdot{\bf
  p}|0\rangle$. Thus, the configuration mixing can make transitions that
are forbidden in the single-particle single-band approximation become allowed
in the many-particle representation of Eq.  (\ref{absorption}) by borrowing
oscillator strength from bright transitions.

%
\begin{figure}
\includegraphics[width=8.5cm]{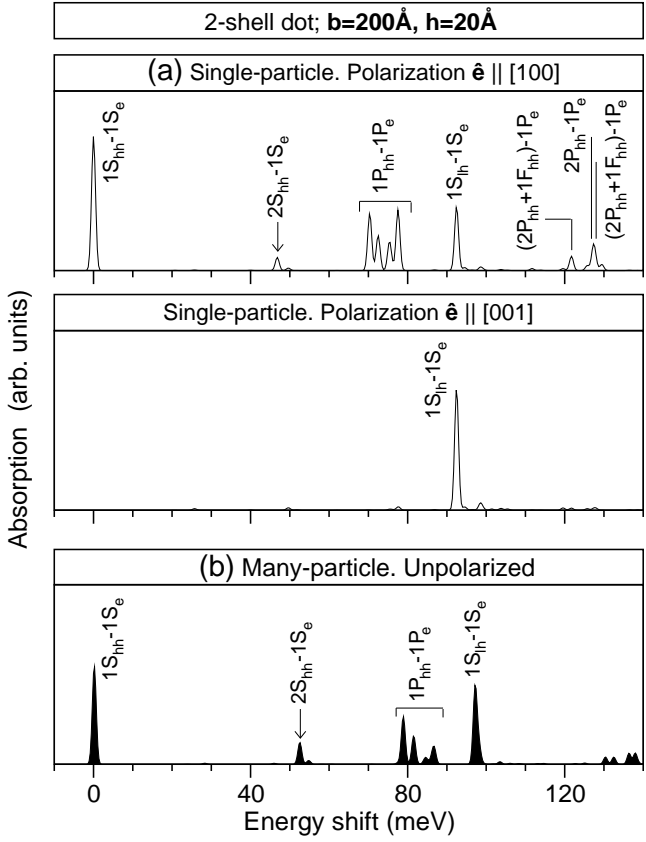}
\caption{{\label{Fig_1}}Optical absorption spectrum of $X^0$ in a 
  lens-shaped In$_{\rm 0.6}$Ga$_{\rm 0.4}$As/GaAs quantum dot (base diameter
  $b=200\;${\AA}, height $h=20\;${\AA}) calculated at the (a) single-particle
  approximation [Eq. (\ref{absorption.SP})] under in-plane ($\hat{\bf
    e}\parallel [100]$; top) and out-of-plane ($\hat{\bf e}\parallel [001]$;
  bottom) polarization; and (b) at the many-particle level [Eq.
  (\ref{absorption})] for unpolarized light. Energy is shown relative to (a)
  the single-particle gap ${\cal E}^{\,(e)}_0-{\cal E}^{\,(h)}_0=1333\;{\rm
    meV}$ and (b) the ground-state energy $E^{(0)}(X^0)=1309\;{\rm meV}$ of
  $X^0$.}
\end{figure}

\subsection{Results}
\begin{figure}
\includegraphics[width=8.5cm]{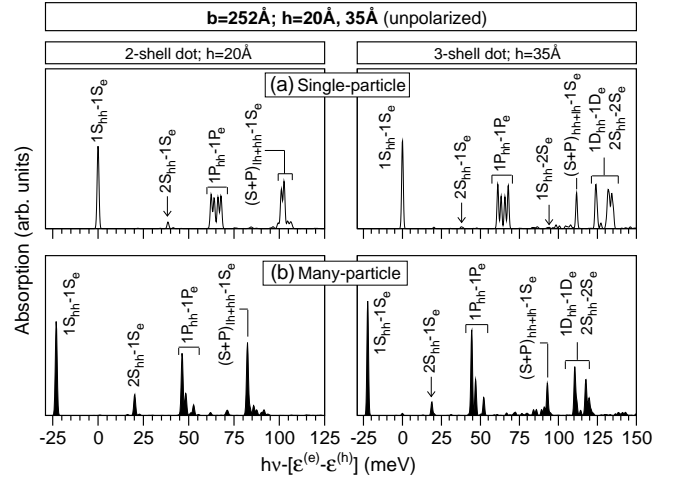}
\caption{{\label{Fig_2.verynew}}{\em Idem} Fig. \ref{Fig_1}(b) for two
  In$_{\rm 0.6}$Ga$_{\rm 0.4}$As/GaAs quantum dots that confine two (left
  panel) and three (right) shells of electron states, with heights
  $h=20\;${\AA} and $30\;${\AA}, respectively, and base $b=252\;${\AA}.}
\end{figure}

%
%
%
%
Figure
\ref{Fig_1} shows our calculated single-particle [Eq.  (\ref{absorption.SP});
Fig.  \ref{Fig_1}(a)] and many-particle [Eq.  (\ref{absorption}); Fig.
\ref{Fig_1}(b)] absorption spectrum of $X^0$ for a lens-shaped In$_{\rm
  0.6}$Ga$_{\rm 0.4}$As/GaAs quantum dot with base diameter $b=200\;${\AA} and
height $h=20\;${\AA} that confines two shells of electron states:
$\{1S_e;1P_e\}$. The energy of the transitions is shown as a shift $\Delta
{\cal E}$ from the single-particle exciton gap ${\cal E}^{(e)}_0-{\cal
  E}^{(h)}_0$ [in Fig. \ref{Fig_1}(a)] or the ground-state energy of the
monoexciton $E^{(0)}(X^0)$ [in Fig.  \ref{Fig_1}(b)]. Figure
\ref{Fig_2.verynew} shows equivalent results for two dots with $b=252\;${\AA}
and heights $h=20\;${\AA} and $35\;${\AA}, which confine two $\{1S_e;1P_e\}$
and three $\{1S_e;1P_e;1D_e+2S_e \}$ shells of electron
states,respectively.\cite{narvaez_JAP_2005} 
As expected, we find nominally-allowed single-particle transitions,
including (i) the fundamental transition $1S_{hh}$-$1S_{e}$ at $\Delta{\cal
  E}=0\;{\rm meV}$; (ii) the $1P_{hh}$-$1P_e$ transitions with energy shifts
$\Delta {\cal E}\sim 75\;{\rm meV}$ and $65\;{\rm meV}$ for dots with
$b=200\;${\AA} and $252\;${\AA}, respectively; and (iii) the transitions
$1D_{hh}$-$1D_e$ and $2S_{hh}$-$2S_e$ at $\Delta {\cal E}\sim 130\;{\rm meV}$
for the three-shell dot ($b=252\;${\AA},$h=35\;${\AA}).
Note that the underlying atomistic $C_{2v}$ symmetry of the circular-base
lens-shape dot splits the electron and hole $1P$ and $1D$ states into 3 and 5
levels, respectively, and causes these states to be a mixture of $L_z=\pm 1$
and $L_z=\pm 2$, respectively. [$L_z$ is the projection of the angular
momentum along the cylindrical ($[001]$, out-of-plane) axis of the dot.]
Thus, in contrast to predictions of simplified models that assume $C_{\infty
  v}$ shape symmetry, transitions involving states $1P$ and $1D$ are split
into four and nine lines, respectively (Figs. \ref{Fig_1} and
\ref{Fig_2.verynew}).
We next discuss the {\em nominally-forbidden} transitions.

%

\subsubsection{Band-offset and orbital-mixing induced 1S-2S transitions}
%
%
%
%
If the electron and hole envelope wave functions are identical, the envelope-function
selection rules indicate that only $\Delta i=0$ ($i\rightarrow i$) transitions
are allowed, as assumed e.g. in Refs.
[\onlinecite{woggon_book,narvaez_PRB_2001,hawrylak_PRL_2000,findeis_SSC_2000,honester_APL_1999}].
This will be the case in the single-band effective mass approximation if the
confinement potential (band offset between dot and environment) for electron
and hole are infinite.
In contrast, we find a few $\Delta i\neq 0$ transitions with significant
intensity: (i) $2S_{hh}$-$1S_e$ [Figs. \ref{Fig_1}(a) and
\ref{Fig_2.verynew}(a)], which we find {\em below} $1P_{hh}$-$1P_e$; (ii) four
transitions that involve the electron states $1P_e$ and hole states $2P_{hh}$
(also found in Ref.  \onlinecite{vasanelli_PRL_2002}) and $2P_{hh}+1F_{hh}$
[Fig. \ref{Fig_1}(a)]; and (iii) transitions $1S_{hh}$-$2S_e$, and
$2S_{hh}$-$1D_e$ and $1D_{hh}$-$2S_e$ in the three-shell dot (Fig.
\ref{Fig_2.verynew}). 
There are two reasons why $\Delta i\neq 0$ transitions are allowed. First, in
the case of {\em finite} band offsets or, equivalently, when the electron and
hole wavefunctions are not identical, the condition $\Delta i=0$ is relaxed
and transitions $j\rightarrow i$ may be allowed even in the effective-mass
approximation. The latter happens to be the case in the work of Vasanelli {\em
  et al.} (Ref.  \onlinecite{vasanelli_PRL_2002}) in which $2S_{hh}$-$1S_e$
and $2P_{hh}$-$1P_e$ transitions between confined electron and hole levels
were found to have finite, non-negligible oscillator strength.
Second, orbital mixing also makes such transition allowed: For example, a dot
made of zinc-blende material and having a lens or cylindrical shape has the
atomistic symmetry $C_{2v}$ while spherical dots have $T_d$ symmetry.  
In contrast, continuum-like effective-mass based theories for dots use
artificially higher symmetries.
In fact, the ability of the envelope-function approximation to recognize the
correct point-group symmetry depends on the number N of $\Gamma$-like bands
used in the expansion.\cite{zunger_pssa_2002} N=1 corresponds to the
``particle-in-a-box'' or to the parabolic single-band effective mass
approximation; N=6 corresponds to including the valence band maximum (VBM)
states only; and N=8 corresponds to considering the VBM states plus the
conduction band minimum. Higher values of N have been also
considered.\cite{14-band_k.p,richard_PRB_2004}
In particular, (a) within the N=1 single-band effective-mass approximation one
uses the symmetry of the {\em macroscopic shape} (lens, cylinder, pyramid,
sphere) rather than the true {\em atomistic} symmetry. For example, for
zinc-blende lenses and cylinders one uses $C_{\infty v}$ symmetry rather than
the correct $C_{2v}$. (b) The 8-band ${\bf k}\cdot{\bf p}$ Hamiltonian assumes
cubic ($O_h$) symmetry to describe the electronic structure of the
dots,\cite{8-band_k.p} the resulting symmetry group is dictated by {\em both}
the symmetry of the macroscopic shape and the cubic symmetry. In the case of
square-pyramid-shaped dot the symmetry is $C_{4v}$ rather than the correct
$C_{2v}$.\cite{note_Td_symmetry}

In single-band effective-mass approaches, transition $i$-$j$ is allowed
as long as the overlap between the respective envelope functions is non-zero.
For example, for spherical quantum dots one expects $S$-$S$ transitions to be
allowed but not $D$-$S$ transitions. Yet, in the true point group symmetry of
the {\em zinc-blende sphere} the highest occupied hole state has mixed $S+D$
symmetry, which renders transition $1S_h$-$1D_e$
allowed,\cite{xia_PRB_1989,fu_PRB_1997} and similarly the ``$2S$-$1S$''
transition is allowed because the $2S_{hh}$ state also contains $1S_{hh}$
character.\cite{note_mixing}

%
\begin{figure}[h]
\includegraphics[width=8.5cm]{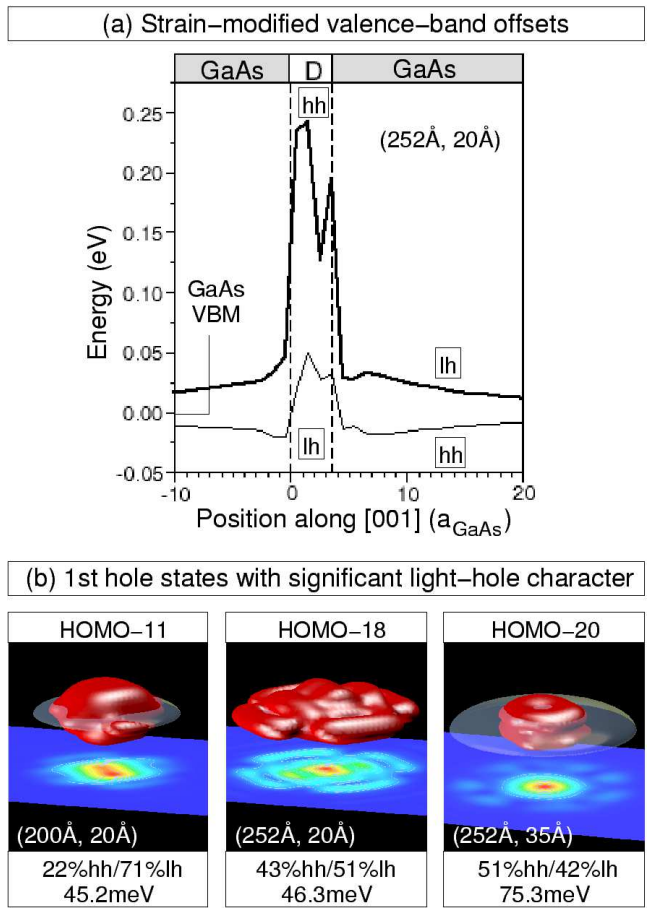}
\caption{{\label{Fig_2.new}} (Color) (a) First (thick line) and second
  (thin line) strain-modified valence-band offsets along a line parallel
  $[001]$ that pierces a lens-shaped In$_{\rm 0.6}$Ga$_{\rm 0.4}$As/GaAs
  quantum dot through its center. The dot size is $b=252\;${\AA} and
  $h=20\;${\AA}. Position is measured in units of the lattice parameter of
  GaAs ($a_{\rm GaAs}$) and the energies are relative to the GaAs VBM
  [$E_v({\rm GaAs})=-5.620\;{\rm eV}$].
  (b) Wave functions of the first hole state with significant $lh$ character
  for different In$_{\rm 0.6}$Ga$_{\rm 0.4}$As/GaAs dots. Isosurfaces enclose
  $75\%$ of the charge density, while contours are taken at $1\;{\rm nm}$
  above the base.  The energy ${\cal E}^{\,(h)}_{j}-E_v({\rm GaAs})$ of the
  state appears in each panel.}
\end{figure}

\subsubsection{Strong light-hole--electron transitions}
%
%
%
%
In {\em bulk} zinc-blende
semiconductors the valence-band maximum is made of degenerate heavy-hole
($hh$, $|3/2,\pm 3/2\rangle$) and light-hole ($lh$, $|3/2,\pm 1/2\rangle$)
states.\cite{bastard_monography} While {\em both} optical transitions
$hh$-$\Gamma_{1c}$ and $lh$-$\Gamma_{1c}$ are polarized in the $\hat{\bf
  x}$-$\hat{\bf y}$ plane, only the latter transition presents polarization
along $\hat{\bf z}$ ($\parallel [001]$). Under biaxial strain these $hh$ and
$lh$ states split. In bulk, the relative energy of these states and their
splitting depend on the strain: for compressive (e.g.  InAs on GaAs) the $lh$
is below the $hh$ while for tensile (e.g. GaAs on InAs) the $lh$ is above the
$hh$.\cite{kent_APL_2002}
%
%
%
In {\em quantum dots} the energy of $lh$ states is unknown. More importantly,
it is generally assumed to be unconfined; so, the $lh$-$\Gamma_{1c}$
transition is expected to be absent from spectroscopic data. Nonetheless,
Minnaert {\em et al}\cite{minnaert_PRB_2001} have speculated that despite the
compressive strain in InAs/GaAs dots the $lh$ state is above the $hh$ states,
while Ribeiro {\em et al}\cite{ribeiro_JAP_2000} have suggested the presence
of a $lh$-derived state {\em below} the $hh$ states by measuring
photo-reflectance and photo-absorption in (In,Ga)As/GaAs dots.
Adler {\em et al}\cite{adler_JAP_1996} and Akimov {\em et
  al}\cite{lh_CdSe/ZnSe} have also suggested the presence of $lh$-derived
transitions in photoluminescence excitation (PLE) experiments in InAs/GaAs and
CdSe/ZnSe self-assembled quantum dots, respectively.
In addition, based on a 6-band ${\bf k}\cdot{\bf p}$ calculation, Tadi\'c {\em
  et al}\cite{tadic_PRB_2002} have predicted that in disk-shaped
InP/In$_{0.51}$Ga$_{0.49}$P dots the light-hole states are confined at the
interface of the disk and become higher in energy than heavy-hole states as
the thickness of the disk is increased.

We show in Fig. \ref{Fig_2.new}(a) the strain-modified valence-band offsets of
a lens-shaped In$_{\rm 0.6}$Ga$_{\rm 0.4}$As/GaAs with $b=252\;${\AA} and
$h=20\;${\AA}, calculated along a line normal to the dot base that pierces the
dot through its center. The energy is presented relative to the GaAs VBM
($E_v(GaAs)=-5.620\;{\rm eV}$). We find that inside the dot the heavy-hole
($hh$) potential is above the light-hole ($lh$) one, while outside this order
is reversed; although the $lh$ character of the lower-energy band-offset leaks
slightly into the barrier close to the dot. Because the dot is alloyed the
band offsets inside the dot are jagged.
In agreement with Ribeiro {\em et al},\cite{ribeiro_JAP_2000} but in contrast
with Minnaert {\em et al},\cite{minnaert_PRB_2001} our atomistic
pseudopotential calculations reveal weakly {\em confined} light-hole states at
energies deeper than the first $hh$ state. The wave functions of
the first of these states are shown in Fig. \ref{Fig_2.new}(b). The energy
spacing between the highest hole state [HOMO ($\psi^{(h)}_0$)] and the deep
$lh$-type states increases from $92.4\;{\rm meV}$ to $101.1\;{\rm meV}$ and
$111.7\;{\rm meV}$ for HOMO-11 [$\psi^{(h)}_{11}$], HOMO-18
[$\psi^{(h)}_{18}$], and HOMO-20 [$\psi^{(h)}_{20}$], respectively.
%
These states give raise to two $lh$-derived transitions in
the absorption spectra: (i) $1S_{lh}$-$1S_e$ [Fig. \ref{Fig_1}] with the deep
$lh$-type state being a mixture of $71\% \;lh$ and $22\% \;hh$.  As seen in
Fig. \ref{Fig_1}, this transition has a large intensity in both $\hat{\bf
  e}\parallel [100]$ and $\hat{\bf e}\parallel [001]$ polarizations.
(ii) $(S+P)_{lh+hh}$-$1S_e$ (Fig. \ref{Fig_2.verynew}) with $hh$/$lh$
percentages of $43$/$51$ and $51$/$42$ for the two-shell and three-shell dots,
respectively. In these dots with $b=252\;${\AA}, the larger base size reduces
the spacing between confined hole states and promotes character mixing. In the
2-shell dot, the offset energy of this transition with respect to
$1P_{hh}$-$1P_e$ is $36.0\;{\rm meV}$, in excellent agreement with the
observed value.\cite{preisler_private}
Note that simple models that follow the common assumption of unconfined $lh$
states do not explain the observed feature.

\subsubsection{Coulomb-induced transitions that are forbidden in the single-particle
  description}
%
%
%
%
%
Due to the electron-hole Coulomb interaction, each
monoexciton state $\Psi (X^0)$ is a mixture of electron-hole configurations
[Eq. (\ref{X0.states})]. This mix results in a enhancement/diminishment of the
intensity of both allowed and nominally-forbidden transitions in the
absorption spectra [Fig.  \ref{Fig_2.verynew}(b)].
These are shown by comparing Fig. \ref{Fig_1}(a) {\em vs} Fig. \ref{Fig_1}(b)
and Fig. \ref{Fig_2.verynew}(a) {\em vs} Fig. \ref{Fig_2.verynew}(b). The
many-body effects include (i) enhancement of the intensity of the nominally
forbidden transition $2S_{hh}$-$1S_{e}$ particularly in the 3-shell dot (Fig.
\ref{Fig_2.verynew}); (ii) redistribution of the intensity of both the
nominally allowed $1P_{hh}$-$1P_{e}$ transitions and the $1D_{hh}$-$1D_e$ and
$2S_{hh}$-$2S_e$ transitions; and (iii) change of the intensity of the
transitions involving deep hole states with significant light-hole character.
The mixing enhancement
$\eta(2S_{hh}-1S_e)=I^{(CI)}(2S_{hh}-1S_e)/I(2S_{hh}-1S_e)=3.2$ and
$\eta(1S_{hh}-1S_e)=1.1$ for the 2-shell dot, while $\eta(2S_{hh}-1S_e)=8.2$
and $\eta(1S_{hh}-1S_e)=1.3$ for the three shell dot. For both dots, the
enhancement of transition $2S_{hh}$-$1S_e$ arises mainly from configuration
mixing with the four configurations $|1P_{hh}1P_{e}\rangle$.
The degree of mixing is {\em small}, $\sim 2\%$ for both dots, due to the {\em
  large} ($\sim 26\;{\rm meV}$) energy splitting between these electron-hole
configurations at the single-particle (non-interacting) level, yet sufficient
to cause a sizeable enhancement of the intensity. We find that the larger
$\eta(2S_{hh}-1S_e)$ for the 3-shell dot arises from a larger mixing with
configuration $|1S_{hh}1S_{e}\rangle$.

%
\begin{figure}[t]
\includegraphics[width=8.5cm]{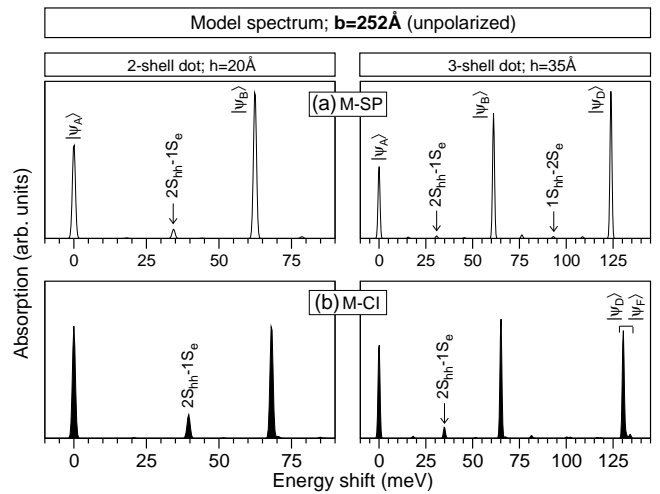}
\caption{{\label{Fig_3}}Optical absorption spectrum of $X^0$ in
  two lens-shaped In$_{\rm 0.6}$Ga$_{\rm 0.4}$As/GaAs quantum dots that
  confine two (left panel) and three (right) shells of electron states.  In
  both cases, the spectra are calculated within a model single-particle (M-SP)
  [Eq. (\ref{absorption.SP})] (a) and configuration-interaction (M-CI)
  approach [Eq. (\ref{absorption})] (b), assuming degenerate $1P$, and $2D$
  and $2S$ (in three-shell dot) electron and hole states but retaining the
  atomistic wavefunctions. Vertical scales are different for each dot.}
\end{figure}

{\em Comparison with experiment:} The calculated $2S_{hh}$-$1S_e$ transition
is $\sim 26\;{\rm meV}$ {\em below} the strongest $1P_{hh}$-$1P_e$ transition;
in excellent agreement with those observed by Preisler {\em et
  al}\cite{preisler_private} in magneto-photoluminescence, who found $25\;{\rm
  meV}$, and in contrast to the effective-mass approximation results of
Vasanelli {\em et al}, which place $2S_{hh}$-$1S_e$ {\em above} $1P_{hh}$-$1P_e$.
The calculated $1S_{lh}$-$1S_e$ transition is $18.3\;{\rm meV}$ below
$1P_{hh}$-$1P_e$, in only rough agreement with the value of $35\;{\rm meV}$
observed by Preisler {\em et al}.\cite{preisler_private}

The effect of configuration-mixing on the optical spectrum was previously
discussed within the simplified single-band 2D-EMA parabolic
model.\cite{narvaez_PRB_2001,hawrylak_PRL_2000} Such continuum theories assume
macroscopic shapes that lead to significant degeneracies among the
single-particle states of Eq.  (\ref{schrodinger}): $P$ states are twofold
degenerate; $D$ states and $2S$ are degenerate; and the $S$-$P$ and $P$-$D$
energy spacings are equal. As a result, there is an artificially strong
many-body mixing in Eq. (\ref{X0.states}).  The many-particle exciton states
with allowed Coulomb mixing are:

\begin{equation}
\label{X.configs}
\begin{array}{rcl}
|\Psi_{A}\rangle &=&|1S_{hh}\,1S_e\rangle \\
|\Psi_{B}\rangle &=&\frac{1}{\sqrt{2}}(|1P^{(+)}_{hh}\,1P^{(+)}_{e}\rangle+|1P^{(-)}_{hh}\,1P^{(-)}_{e}\rangle) \\
|\Psi_{H}\rangle &=&\frac{1}{\sqrt{2}}(|2S_{hh}\,1S_e\rangle+|1S_{hh}\,2S_e\rangle) \\
|\Psi_{D}\rangle &=&\frac{1}{\sqrt{3}}(|1D^{(+)}_{hh}\,1D^{(+)}_{e}\rangle+|1D^{(-)}_{hh}\,1D^{(-)}_{e}\rangle+|2S_{hh}\,2S_e\rangle) \\
|\Psi_{F}\rangle &=&\frac{1}{\sqrt{6}}(|1D^{(+)}_{hh}\,1D^{(+)}_{e}\rangle+|1D^{(-)}_{hh}\,1D^{(-)}_{e}\rangle-2|2S_{hh}\,2S_e\rangle).
\end{array}
\end{equation}

\noindent Here, the $(\pm)$ labels indicate $L_z=\pm 1$ and $\pm 2$ for the 
$P$ and $D$ states, respectively.
The Coulomb interaction couples states $|\Psi_B\rangle$ and $|\Psi_H\rangle$.
Thus the $P$-$P$ transition is split in two lines $\Psi^{(s,p)}_a \simeq
|\Psi_B\rangle+|\Psi_H\rangle$ and $\Psi^{(s,p)}_b \simeq
|\Psi_B\rangle-|\Psi_H\rangle$. States $|\Psi_D\rangle$ and $|\Psi_F\rangle$,
which arise from nominally allowed electron-hole configurations, are also
coupled; consequently, the $D$-$D$ transition splits in two lines
$\Psi^{(d,d)}_a \simeq |\Psi_D\rangle+|\Psi_F\rangle$ and $\Psi^{(d,d)}_b
\simeq |\Psi_D\rangle-|\Psi_F\rangle$. 
The mixing enhancement $\eta(|\psi_H\rangle)$ within this model is
$\infty$ (because transitions $2S_{hh}$-$1S_e$ and $1S_{hh}$-$2S_e$ are {\em
  forbidden} at the single-particle level).

To compare our atomistic predictions with the model calculations, we {\em
  deliberately neglect} in the pseudopotential-based calculation the
atomistic-induced splitting of the $1P$, and $1D$ and $2S$ states (but
preserve their atomistically calculated wavefunctions). We calculate the
absorption spectrum at the single-particle level [Eq. (\ref{absorption.SP})]
and separately in the many-particle approximation [Eq.
(\ref{absorption})].\cite{note.00,note_many-body} 
Figures \ref{Fig_3}(a) and \ref{Fig_3}(b) show, respectively, the atomistic
model calculation of the {\em single-particle} and {\em many-particle}
absorption spectra. By comparing the results of Fig.  \ref{Fig_3} (atomistic
wavefunctions; no $P$ or $D$ splittings) with the expectations from Eq.
(\ref{X.configs}) (continuum wavefunctions; no $P$ or $D$ splittings), we find
that
(i) in the atomistic calculation the CI-enhanced transition corresponds to a
mixture of states $|2S_{hh}\,1S_e\rangle$ and $|\Psi_B\rangle$ instead of a
mixture of $|\psi_H\rangle$ and $|\psi_B\rangle$ as in the model of Eq.
(\ref{X.configs}); and (ii) the $D$-shell transition peak [$|\psi_D\rangle$,
Fig.  \ref{Fig_3}(a)] splits in two transitions that correspond to a mixture
of $|\psi_D\rangle$ and $|\psi_F\rangle$ as in the 2D-EMA model.

\subsection{Comparison with 8-band ${\bf k}\cdot{\bf p}$ calculations of the interband optical spectrum}

Other authors have calculated the absorption spectrum of pure, non-alloyed
InAs/GaAs quantum dots using the 8-band ${\bf k}\cdot{\bf p}$ method with
cubic symmetry. A comparison with our atomistic, pseudopotential-based
predictions in alloyed (In,Ga)As/GaAs dots shows the following main features.

(i) Our prediction of transition $2S_{hh}$-$1S_{e}$ between $1S$-$1S$ and
$1P$-$1P$ is consistent with the findings of nominally-forbidden transitions
between $1S$-$1S$ and $1P$-$1P$ by Heitz {\em et al.}\cite{heitz_PRB_2000},
who calculated the (many-body) absorption spectrum of a monoexciton in
pyramid-shaped non-alloyed InAs/GaAs dots with base length of $d=170\;${\AA}
(height unspecified); Guffarth {\em et al.},\cite{guffarth_PRB_2003} who
calculated the (many-body) absorption spectra of truncated-pyramid InAs/GaAs
dots ($d=180\;${\AA},$h=35\;${\AA}); and the single-particle calculations of
Sheng and Leburton\cite{sheng_PSSB_2003} in the case of a pure non-alloyed
lens-shaped InAs/GaAs dot with $d=153\;${\AA} and $h=34\;${\AA}.
Conversely, other 8-band ${\bf k}\cdot{\bf p}$ plus CI calculations did not
predict nominally-forbidden transitions between $1S$-$1S$ and $1P$-$1P$, like
those of Stier {\em et al.}\cite{stier_PSSA_2002} for a truncated-pyramid
InAs/GaAs dot with height $h=34\;${\AA} (base lenght unspecified), who found
three groups of transitions: $1S$-$1S$, $1P$-$1P$, and $1D$-$1D$, {\em
  without} the presence of satellites around transitions $1P$-$1P$.
Similarly, recent calculations by Heitz {\em et al.}\cite{heitz_PRB_2005} of
the absorption spectrum for small, flat ($d=136\;${\AA} and heights from
$3$-$7\;${\rm ML}) truncated-pyramid InAs/GaAs dots also predicted the {\em
  absence} of nominally-forbidden transitions between $1S$-$1S$ and $1P$-$1P$.

(ii) We predict that transitions $1P$-$1P$ and $1D$-$1D$ are split and span
about $10\;{\rm meV}$ and $15\;{\rm meV}$ respectively.  Instead, the ${\bf
  k}\cdot{\bf p}$-based calculations of Heitz {\em et
  al.}\cite{heitz_PRB_2000} predict that $1P$-$1P$ and 1D-1D transitions are
much heavily split, each group spanning about $50\;{\rm meV}$.

(iii) Sheng and Leburton\cite{sheng_APL_2002} calculated the single-particle
dipole oscillator strength for a truncated-pyramid InAs/GaAs dot with
$d=174\;${\AA} and $h=36\;${\AA} and found strong nominally-forbidden
transitions $1D$-$1P$ nearly $50\;{\rm meV}$ above transitions $1P$-$1P$, and
a transition HOMO-7-to-$2S$.  
Our predictions differ from these in that we find transitions
$(2P_{hh}+1F_{hh})$-$1P_e$ above $1P$-$1P$ [Fig. \ref{Fig_1}(a)]. In addition,
in this energy interval ($\sim 50\;{\rm meV}$ above $1P$-$1P$) we do not
predict hole states with nodes along the [001] axis of the dots.

(iv) None of the 8-band ${\bf k}\cdot{\bf p}$-based plus CI calculations of
Refs. \onlinecite{heitz_PRB_2000,guffarth_PRB_2003,stier_PSSA_2002,heitz_PRB_2005}
predicted strong light-hole--to--conduction transitions originating from deep,
weakly-confined hole states with predominant $lh$ character lying between the
$1P$-$1P$ and $1D$-$1D$ transitions.

\section{Conclusion}
%
%
%
Atomistic, pseudopotential-based calculations of the excitonic absorption of
lens-shaped (In,Ga)As/GaAs quantum dots predict nontrivial spectra that show
nominally-forbidden transitions allowed by single-particle band-offset effects
as well as enhanced by many-body effects, and transitions involving deep,
weakly confined hole states with significant light-hole character. These
transitions explain the satellites of the $P$-$P$ nominally-allowed
transitions recently observed in PLE.

\begin{acknowledgments}

The authors thank G. Bester and L. He for valuable discussions, and V.
Preisler (ENS, Paris) for making data in Ref. \onlinecite{preisler_private}
available to them prior to publication.  This work was funded by U.S.
DOE-SC-BES-DMS, under Contract No.  DE-AC3699GO10337 to NREL.

\end{acknowledgments}


\begin{thebibliography}{100}


\bibitem{heitz_PRB_2000} R. Heitz, O. Stier, I. Mukhametzhanov, A. Madhukar,
  and D. Bimberg, Phys. Rev. B {\bf 62}, 11017 (2000).

\bibitem{guffarth_PRB_2003} F. Guffarth, R. Heitz, A. Schliwa, O. Stier,
  M. Geller, C. M. A. Kapteyn, R. Sellin, and D. Bimberg, Phys. Rev. B {\bf
  67}, 235304 (2003).

\bibitem{sheng_PSSB_2003} W. Sheng and J.-P. Leburton, phys. stat. sol. (b)
  {\bf 237}, 394 (2003).

\bibitem{stier_PSSA_2002} O. Stier, R. Heitz, A. Schliwa, and D. Bimberg,
  phys. stat. sol. (a) {\bf 190}, 477 (2002).

\bibitem{heitz_PRB_2005} R. Heitz, F. Guffarth, K. P\"otschke, A. Schliwa,
  D. Bimberg, N. D. Zakharov, and P. Werner, Phys. Rev. B {\bf 71}, 045325 (2005).

\bibitem{sheng_APL_2002} W. Sheng and J.-P. Leburton, Appl. Phys. Lett. {\bf
    80}, 2755 (2002).


\bibitem{zunger_pssb_2001} A. Zunger, phys. stat. sol. (b) {\bf 224 }, 727 (2001).

\bibitem{wang_PRB_1999} L.-W. Wang and A. Zunger, Phys. Rev. B {\bf 59}, 15806 (1999).

\bibitem{williamson_PRB_2000} A. J. Williamson, L. W. Wang, and A. Zunger, Phys. Rev. B {\bf 62}, 12963 (2000).

\bibitem{pryor_JAP_1998} C. Pryor, J. Kim, L. W. Wang, A. J. Williamson, and
  A. Zunger, J. Appl. Phys. {\bf 83}, 2548 (1998).


\bibitem{kim_PRB_2002} K. Kim, P. R. C. Kent, A. Zunger, and C. B. Geller,
  Phys. Rev. B {\bf 66}, 045208 (2002) and references therein.



\bibitem{cardona_book}  P. Y. Yu and M. Cardona, 
{\em Fundamentals of Semiconductors---Physics and Materials Properties} (Springer, Berlin, 2003).

\bibitem{franceschetti_PRB_1999} A. Franceschetti, H. Fu, L. W. Wang, and A. Zunger, Phys. Rev. B {\bf 60}, 1819 (1999).

\bibitem{note_unpolarized} Unpolarized photons incoming from all directions.

\bibitem{narvaez_JAP_2005} G. A. Narvaez, G. Bester, and A. Zunger, J. Appl. Phys. {\bf 98}, 043708 (2005).
 
\bibitem{woggon_book} U. Woggon, {\em Optical properties of semiconductor
    quantum dots}, (Springer-Verlag, Berlin, 1997).

\bibitem{narvaez_PRB_2001} G. A. Narvaez and P. Hawrylak, Phys. Rev. B {\bf
    61}, 13753 (2001).

\bibitem{hawrylak_PRL_2000} P. Hawrylak, G. A. Narvaez, M. Bayer, and A. Forchel, Phys. Rev. Lett. {\bf 85}, 389 (2000).

\bibitem{findeis_SSC_2000} F. Findeis, A. Zrenner, G. B\"ohm, and G. Abstreiter, Solid State Comm. {\bf 114}, 227 (2000).

\bibitem{honester_APL_1999} U. Honester, R. Di Felice, E. Molinari, and F. Rossi, Appl. Phys. Lett. {\bf 75}, 3449 (1999).

\bibitem{vasanelli_PRL_2002} A. Vasanelli, R. Ferreira, and G. Bastard, Phys. Rev. Lett. {\bf 89}, 216804 (2002).

\bibitem{zunger_pssa_2002} A. Zunger, phys. stat. sol. (a) {\bf 190}, 467 (2002).

\bibitem{14-band_k.p} N. Rougemaille, H.-J. Drouhin, S. Richard, G. Fishman, and A. K. Schmid, Phys. Rev. Lett. {\bf 95}, 186406 (2005);
C. W. Cheah, L. S. Tan, G. Karunasiri, J. Appl. Phys. {\bf 91}, 5105 (2002);
W. H. Lau, J. T. Olesberg, and M. E. Flatt\'e, Phys. Rev. B {\bf 64},
161301(R).

\bibitem{richard_PRB_2004} S. Richard, F. Aniel, and G. Fishman, Phys. Rev. B
  {\bf 70}, 235204 (2004); Phys. Rev. B {\bf 71}, 169901(E) (2005).

\bibitem{8-band_k.p} C. Pryor, Phys. Rev. B {\bf 57}, 7190 (1998); W. Sheng
  and J.-P. Leburton, Phys. Stat. Sol. (b) {\bf 237}, 394 (2003).
  
\bibitem{note_Td_symmetry} It is possible to have an 8-band ${\bf k}\cdot{\bf
    p}$ Hamiltonian for strained zinc-blende semiconductors with the correct
  tetrahedral ($T_d$) symmetry. [See H.-R. Trebin, U. R\"ossler, and R.
  Ranvaud, Phys. Rev.  B {\bf 20}, 686 (1979).] If this Hamiltonian is used to
  find the electronic energy-level structure of a square-pyramid-shaped
  InAs/GaAs dot the symmetry of the problem would reduce to $C_{2v}$.


\bibitem{xia_PRB_1989} J.-B. Xia, Phys. Rev. B {\bf 40}, 8500 (1989).

\bibitem{fu_PRB_1997} H. Fu and A. Zunger, Phys. Rev. B {\bf 56}, 1496 (1997).
  
\bibitem{note_mixing} Band-mixing effects mix the $1S$ and $2S$ character of
  the single-particle hole states.

\bibitem{bastard_monography} G. Bastard, {\em Wave Mechanics Applied to
    Semiconductor Heterostructures} (Halstead, New York, 1988).

\bibitem{kent_APL_2002} P. R. C. Kent, G. L. W. Hart, and A. Zunger, Appl. Phys. Lett. {\bf 81}, 4377 (2002).
  
\bibitem{minnaert_PRB_2001} A. W. E. Minnaert, A. Yu. Silov, W. van der
  Vleuten, J. E. M. Haverkort, and J. H. Wolter, Phys. Rev. B {\bf 63}, 075303 (2001).
  
\bibitem{ribeiro_JAP_2000} E. Ribeiro, F. Cerdeira, M. J. S. P. Brasil, T.
  Heinzel, K. Ensslin, G. Medeiros-Ribeiro, and P. M. Petroff, J. Appl. Phys.
  {\bf 87}, 7994 (2000).
  
\bibitem{adler_JAP_1996} F. Adler, M. Geiger, A. Bauknetch, F. Scholz, H.
  Schweizer, M. H. Pilkuhn, B. Ohnesorge, and A. Forchel, J. Appl. Phys. {\bf
    80}, 4019 (1996).
  
\bibitem{lh_CdSe/ZnSe} A. I. Akimov, A. Hundt, T. Flissikowski, P. Kratzert
  and F. Henneberger, Physica E {\bf 17}, 31 (2003); T. Flissikowski, I. A.
  Akimov, A. Hundt, and F. Henneberger, Phys.  Rev. B {\bf 68}, 161309(R)
  (2003).

\bibitem{tadic_PRB_2002} M. Tadi\'c, F. M. Peeters, and K. L. Janssens,
  Phys. Rev. B {\bf 65}, 165333 (2002).

  
\bibitem{preisler_private} V. Preisler, T. Grange, R. Ferreira, L. A. de
  Vaulchier, Y. Guldner, F. J. Teran, M. Potemski, and A. Lema$\hat{\rm
    i}$tre, Phys. Rev. B {\bf 73}, 075320 (2006).

\bibitem{note.00} Both dots confine more than three shells of hole states but
  we only consider three in this model calculation.
  
\bibitem{note_many-body} In the pseudopotential-based plus CI calculation we
  do not use Eq.  (\ref{X.configs}), but we let the Coulomb interaction mix
  the configurations $|e_ih_j\rangle$ that compose the monoexciton states [Eq.
  (\ref{X0.states})]. We include the electron-hole direct Coulomb
  coupling but neglect the much smaller electron-hole exchange.


\end{thebibliography}
\end{document}